\documentclass[twocolumn,preprintnumbers,amsmath,amssymb]{revtex4}

\usepackage{graphicx}  
\begin{document}

\title
{Infrared photodetectors based on graphene van der Waals  heterostructures}

\author{V~Ryzhii$^{1,2,3}$, M~Ryzhii$^4$,
D~Svintsov$^5$, V~Leiman$^5$,  V~Mitin$^{1,6}$, M~S~Shur$^7$
 and T~Otsuji$^1$}
\address{
$^1$ Research Institute of Electrical Communication, Tohoku University,
 Sendai 980-8577, Japan\\
$^2$ Institute of Ultra High Frequency Semiconductor Electronics of RAS,\\
 Moscow 117105, Russia\\
$^3$ Center for Photonics and Infrared Engineering, Bauman Moscow State Technical University, Moscow 111005, Russia\\
$^4$ Department of Computer Science and Engineering, University of Aizu, 
Aizu-Wakamatsu 965-8580, Japan\\
$^5$ Laboratory of 2D Materials' Optoelectronics, Moscow Institute of Physics and Technology, Dolgoprudny 141700, Russia\\
$^6$ Department of Electrical Engineering, University at Buffalo,\\ Buffalo, New York 1460-1920, USA\\
$^7$ Department of Electrical, Computer, and Systems Engineering, Rensselaer Polytechnic Institute, Troy, New York 12180, USA}
%


\begin{abstract}
We propose and evaluate the graphene layer (GL)  infrared photodetectors (GLIPs)   based on the
van der Waals (vdW)  heterostructures with the radiation absorbing GLs.
The operation  of the GLIPs  is associated with the electron photoexcitation 
from the GL valence band to the continuum states above the inter-GL barriers 
(either via tunneling  or  direct  transitions to the 
continuum states).  Using the developed device model, we calculate the photodetector characteristics as functions of the GL-vdW heterostructure parameters. 
We show that due to
a relatively large efficiency of the electron photoexcitation and low capture efficiency of the electrons propagating over the barriers in the inter-GL layers, GLIPs should exhibit the
elevated photoelectric gain and  detector responsivity as well as relatively high detectivity.
The possibility of high-speed operation, high conductivity, transparency of the GLIP contact layers, and the sensitivity to normally incident IR radiation
 provides additional potential advantages in comparison with other IR photodetectors.
In particular, the proposed GLIPs can compete with unitravelling-carrier photodetectors.
\end{abstract}

\maketitle

%
%
%
%
%


%

\section{Introduction}

Successful development  of the van der Waals (vdW) heterostructures~\cite{1} based on stacking of two-dimensional (2D)
crystals and graphene layers (GLs) promises a significant progress in infrared (IR) 
and terahertz (THz) optoelectronics
~\cite{2,3}. The unique  properties of GLs
  provide an opportunity to
create new effective photodetectors operating in a wide range of photon energies~
\cite{4,5,6,7,8,9,10,11,12,13}. Supplementing their design by the band structure engineering using 2D-materials for the barrier layers opens up new prospects in  further enhancement of the detector performance.
Recently, new IR and  THz detectors using  GL-vdW heterostructures were proposed and 
evaluated:\\
(i) THz photodetectors based on a double-GL structure, in which the {\it intraband} inter-GL
resonant transitions are assisted by IR or THz photons~\cite{14}. The resonant nature of such transitions, in which the electron momentum is conserved~\cite{15,16,17,18,19}, promotes their elevated probability
and, hence, the elevated quantum efficiency. These photodetectors require inclined radiation incidence or a radiation coupler (grating structure); \\  
(ii) THz and IR photodetectors based on the GL-vdW heterostructures using the photon-assisted  {\it interband}
transitions between the neighboring GLs~\cite{20}; \\
(iii) THz detectors based on the GL-vdW heterostructures operating as hot-electron bolometers using  the effect of the emission from GLs of the electrons, heated  by the THz
radiation, due to the  {\it intraband} (Drude) absorptions in GLs~\cite{21}.

\begin{figure*}[t]
\centering
\includegraphics[width=15.0cm]{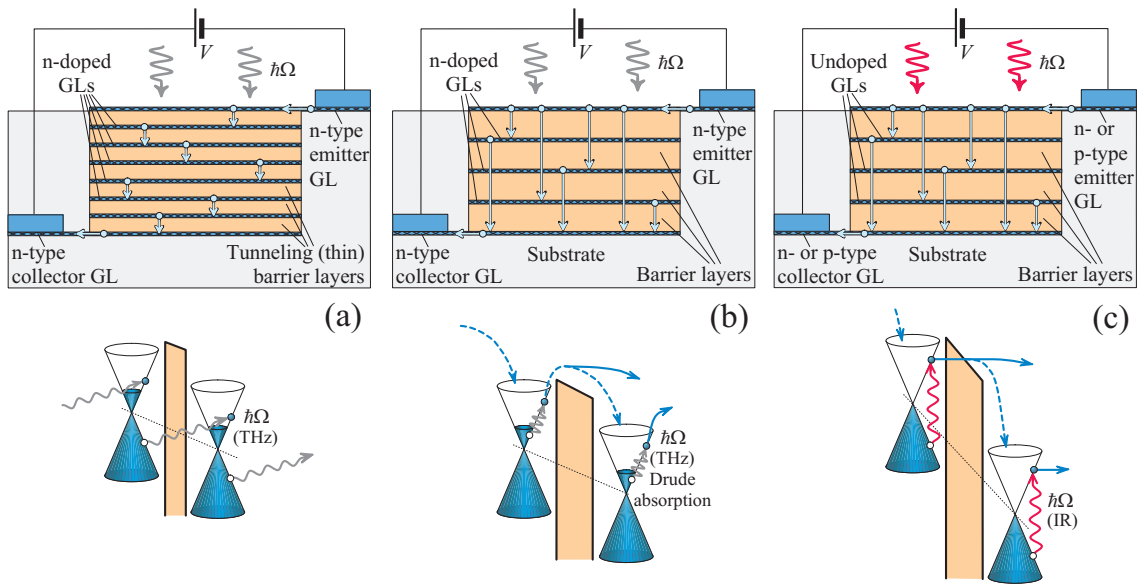}
\caption{Schematic view of the GLIPs 
 based on GL-vdW heterostructures
 (upper panels) with (a)  interband tunneling inter-GL bound-to-bound tunneling transitions (narrow barrier layers)~\cite{20}, (b)  the thermionic emission from GLs of the  electrons 
heated by  THz radiation due to intraband (Drude) absorption~\cite{21} and 
 (c)  interband  bound-to-continuum tunneling-assisted  transitions
of the photoexcited electrons considered in this work. Lower panels show the band diagrams fragments in which the  pertinent radiative processes with the photon absorption (wavy arrows) as well as the escape and capture processes (fine solid and dashed arrows) are indicated.
}
\end{figure*}

In this paper, we propose and evaluate the IR photodetectors based on the GL-vdW heterostructures with different numbers of the active undoped GLs placed between the emitter and collector and separated by the inter-GL barriers (GLIPs). These photodetectors are using the {\it interband} photoexcitation of the electrons in GLs from the valence band states to the exited states slightly below or higher than the barrier edge.
The photoexcited electrons transfer (by tunneling through the barrier top or directly)  to the continuum states above the barrier and propagate through the barrier layers.
The GL-vdW double-GL photoconductive device with the emitter and collector GLs separated by  the barrier and using a similar mechanism  was recently proposed~\cite{22}
fabricated and measured~\cite{23}. 
However, as we demonstrate below, the insertion of the inner GLs, which  provides the extra electron 
photoemission and capture, can lead to a dramatic increase in the GLIP responsivity.
This is 
due to the   photocurrent gain associated with the redistribution
of the electric field in the heterostructure and  leading  to amplification of the electron injection from the emitter.
Thus the GLIPs 
in some sense  resemble quantum-well infrared photodetectors (QWIPs) typically using  III-V materials and  extensively studied, fabricated, and used in applications  during two decades~\cite{24,25,26,27}. 
Such a similarity is  due to the propagation of the photoexcited electrons over the barriers
with their fraction being  captured  back to the bound states (in GLs and QWs, respectively). A relatively low capture probability results in the  elevated values of the detector responsivity. Apart from this, in both GLIPs and QWIPs the detector detectivity increases with
an increasing number of GLs and QWs (see, for example,~\cite{24}). 

The main difference between  the GLIPs and QWIPs is in the use of the interband (in GLIPs) and intraband (in QWIPs) transitions. This allows  eliminating the grating structures required in the n-QWIPs for radiation  coupling. 
In realistic GLIPs with a relatively large conduction band off-set, the proper electron photoexcitation  requires higher photon energies $\hbar\Omega$ than in QWIPs.
This implies that the GLIPs might generally operate at
 shorter radiation wavelengths than the QWIPs. 
However, modifying the shape of the barriers by doping of the inter-GL layers, one can achieve an effective electron photoescape at fairly low photon energies (in far-IR range).
As shown in~\cite{28,29}, the probability of the electron capture into the GLs 
can be much smaller than that for the  QWs~(see, for example,~\cite{30}) since 
doping of the QWs in  the QWIPs  leads to the inclusion of the electron-electron scattering
mechanism of the capture. 
Smaller capture efficiency is beneficial for a higher detector responsivity~\cite{30}
(see also,~\cite{24,25,26,27}). 

The absence of the GL doping promotes lower electron emission
of the thermalized electrons than that in the doped QWs and, hence, leads to a smaller dark current.
Finally, the probability of the electron photoexcitation from the  GLs is substantially higher than from the  QWs. 
These  GLIP features  (the sensitivity to the normally incident radiation and  higher photoexcitation and lower capture probabilities)
 provide potential advantages of the GLIPs over the QWIPs and some other photodetectors. 
The photodetectors based on GL-vdW heterostructures with the tunneling transparent inter-GL barrier layers considered recently~\cite{20} and the GLIPs with relatively thick layers
studied in the present paper are analogous. However their operation principles are quite different:
the photo-stimulated cascade
tunneling between GLs in ~\cite{20} and the
photoexcitation of electrons to the continuum states and propagation in GLIPs. In the latter device, only a small fraction
of the propagating electrons can be  captured to the GLs.
Such a distinction results in different spectral properties and other detector characteristics.
In particular, in the GLIPs under consideration, 
the photoexcitation probability 
can be much larger than the probability of the inter-GL photo-stimulated transfer.
This is 
due to the diminished wave function
overlap in the neighboring GLs~\cite{20}. Apart from this, 
the GLIPs considered here and the hot-electron GLIPs using the bolometric mechanism~\cite{21}
are characterized by a rather low
 capture probability of electrons in the continuum states~\cite{28,29}. 
 This promotes high values of the detector responsivity~\cite{30,31}
This is in contrast with  the detection mechanism  associated  with the cascade inter-GL photo-stimulated tunneling.
The spectral ranges  of the photodetectors using the inter-GL photo-stimulated cascade transfer~{20} and the electron heating~\cite{21} from one side and the spectral range of the GLIP operation can be essentially different. 
One important feature of the GLIPs is that their operation do not require the mutual alignment of the GLs.

Various  device models used for studies of the vertical electron transport
in QWs (and GLs) introduce the concept of an "ideal" emitter (which provides as many injected electrons as the bulk of the structure "requires"). In the framework of this concept, the electric-field distributions are assumed to be uniform. A more accurate consideration
of the transport phenomena  in such structures (including GLIPs) in the dark conditions and under irradiation accounts for   the self-consistent
electric field distributions and the nonideality of the emitter~\cite{32,33,34,35,36,37,38}.
This paper uses this more accurate approach.

The paper is organized as follows. In Section~2, we describe the possible GLIP device structures and the GLIP operation principle. Section~3 deals with  the device mathematical model. The latter includes the equation
for the self-consistent electric potential and the equations governing the electron balance in the GLs.
In the latter equation, the electron capture into the GLs is described phenomenologically invoking the concept of the capture parameter~\cite{30} (see, also,~\cite{26,27,28,29,30}.
In Section~4, we derive  the dark current in the  GLIPs 
 as a function of the structural parameters and the applied voltage using the model described in Sec.~3.
In Section~5, we calculate the photocurrent (the variation of the current under the  incident radiation). 
The expressions obtained in Sections~4 and 5 are then used in Sec.~6 for the derivation of the GLIP detector responsivity, photoelectric gain, and detectivity. 
In Section~6, we compare the GLIPs with some other IR photodetectors.
In the Conclusion section, we draw the main results of the paper. 
In the Appendix, we discuss some simplifications of the main model.

\begin{figure}[t]
\centering
\includegraphics[width=7.0cm]{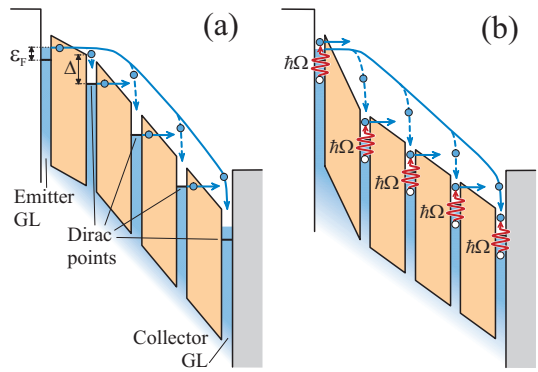}
\caption{Band diagrams of a GLIP under consideration (a) in dark and (b) under strong irradiation.
Bold arrows indicate the tunneling injection of thermalized and photoexcited electrons from the GLs and their propagation in the continuum states above the barriers,
wavy arrows show the electron interband intra-GL photoexcitation with the generation of electron-hole pairs
followed by the electron tunneling escape from the GLs,  dashed arrows correspond to the electron capture  from the continuum states into the GLs .}
\end{figure}

\section{Device structure and operation principle}

We consider the GLIPs based on the GL heterostructure  which consists of the $N$ inner  GLs clad between
the $M = N + 1$ barrier layers with the top and bottom contact GLs.
The conduction band offsets between the emitter GL and the barrier layer, $\Delta_{E}$, as well as between the inner GLs and
the barrier layers between them, $\Delta$, are smaller than  the pertinent offsets in the valence band. Generally, the materials of the emitter barrier layer and
of other barrier layers can be different, so that $\Delta_{E} \neq \Delta$. 

The top and bottom  GLs   are assumed to be doped to provide their sufficiently high lateral conductivity.
These extreme GLs serve as the tunneling emitter (injecting electrons into the GL-heterostructure bulk) and the   collector, respectively.
The top and bottom GLs 
  are supplied with  contacts between which the bias voltage $V$ is applied. 
We consider
the GL heterostructures with either the n-type or p-type doping of both the emitter and collector GLs and undoped
inner GLs and the barrier layers, i.e., the
 heterostructures with different  types of the emitter and collector can also be treated in the framework of our model.
The applied voltage in such  heterostructures, stimulates  the electron tunneling through the barriers with  triangular tops.
The  thermalized electrons tunneling from the GLs and the electrons excited by the incident IR radiation  propagate over the barriers.

Figures 1(a) and 1(b) show schematically the GLIPs based on  GL-vdW heterostructures 
considered in~\cite{20,21} and   their band diagrams.
Figure 1(c) corresponds to the  GLIPs  with the    bound-to-continuum tunneling and photon-assisted  transitions
considered in the present work.

Figure~2 shows the  band diagrams of the GLIP (with the structure of figure~1(c)) using the interband intra-GL photoexcitation
with the sequential tunneling of the photoexcited electrons 
 (a) in  dark conditions and (b) under the strong IR irradiation at the sufficiently high bias voltage ($V > T/e$, where $e = |e|$ is the electron charge and $T$ is the temperature in the energy units).

The applied voltage induces the extra carriers of the opposite signs  and changes the electron Fermi energies in  the emitter and collector GLs.
The electron Fermi energies, $\varepsilon_{F,E}$ and $\varepsilon_{F,C}$, in these GLs (at their equal doping)
are  different: $\varepsilon_{F,E} = \varepsilon_F + \Delta \varepsilon_{F,E}$ and  $\varepsilon_{F,C} = \varepsilon_F + \Delta \varepsilon_{F,C}$
Here $\varepsilon_F \simeq \pm\hbar\,v_W\sqrt{\pi\Sigma_0}$ is the Fermi energy in the extreme GLs
 at their donor (upper sign for  the n-type contact GLs) or acceptor (for the p-type contact GLs) density equal to  $\Sigma_0$, $\hbar$ is the Planck constant,
$v_W \simeq 10^8$~cm/s is the electron and hole velocity in GLs, and $\Delta \varepsilon_{F,E}$ and
$\Delta \varepsilon_{C,F}$.
Hence, the bias voltage can to some extent affect the emitter 
injection characteristics decreasing the electron activation energy in the emitter $\Delta_E
 - \varepsilon_F$ (see the Appendix).

To enhance the electron  collection by the collector, the bottom GL 
can be replaced by a doped multiple-GL or by a thick wider band-gap doped collector region.
In principle, the GLIPs   with the n-type emitters can be supplied with p-type collectors
and vice versa.

The operation of the GLIPs with the device structure  and the band diagrams corresponding to figures~1(c)  
and 2 considered below is associated with the following.
The incoming IR photons generate the electron-hole pairs (due to the photon absorption
and the vertical interband transitions) in each GL. Depending on the ratio 
$[(\hbar\Omega/2 - \Delta)/\Delta]$,  the photogenerated electrons appear in the continuum states above the barriers 
directly (if $\hbar\Omega > 2\Delta$) or via the intermediate excited state with the energy $\varepsilon = \hbar\Omega/2$ with respect to the Dirac point  following the  electron tunneling through the triangular barrier top~\cite{39}  [see figure~2(b)].
Upon the photoexcitation  to the states over the barriers, the  electrons  propagate  
 toward the collector. During the electron transport over the barrier, a fraction of the propagating electrons is captured into the GLs.

Since we consider devices with the barriers, in which the conduction band off-set $\Delta$ and   the quantity $\hbar\Omega/2$ are smaller than the valence band off-set,   the  escape of the photogenerated holes   from the GLs can be neglected. The GLIP barriers in the devices under consideration can be made of the van der Waals structures with the InSe,
WS$_2$, and other barrier layer materials~\cite{1,2,16,39}.

\section{Equations of the model}

The Poisson equation governing the electric potential distribution $\varphi = \varphi (z)$ in the direction perpendicular to the GL plane (the $z-$direction) and the equations governing the electron balance in each GL 
at the steady-state irradiation  are presented in the following form:
 
\begin{equation}\label{eq1}
\frac{d^2\varphi}{d z^2} = \frac{4\pi\,e}{\kappa}\sum_{n = 1}^{N}(\Sigma_n^{e} - \Sigma_n^{h})
\cdot\delta (z - nd), 
\end{equation}

\begin{equation}\label{eq2}
\frac{jp_n}{e} = G_n +  \beta\,\theta_nI.
\end{equation}
Here  $\Sigma_n^{e}$ and $\Sigma_n^{h}$
  are  the 2D densities of electrons and holes 
in the inner GLs with the index $n = 1,2,...,N$ ($n=0$ and $n =N+1 = M$ corresponding to the emitter and collector GLs, respectively),  
$j$ is the density of the electron current across GLs,  $p_n$ is the capture efficiency for the electrons crossing the $n-$GL ~\cite{25,26,27,28}, $G_n$  and $\beta\,\theta_nI_n$ are
the rates of the tunneling emission of the thermalized electrons and  the tunneling emission of the electrons photoexcited (photogenerated) from the GL valence band
in the $n-$th GL, 
 $\beta = \pi\alpha/\sqrt{\epsilon}$ is the probability of the interband photon absorption in GLs, 
 ($\alpha =e^2/\hbar\,c \simeq 1/137$ is the fine structure constant, where $c$ is the speed of light in vacuum and $\sqrt{\epsilon}$ is the barrier material refraction index), 
 $\theta_n$  is the probability of the photoexcited electrons tunneling escape from the $n-$th GL to the continuum states
above the inter-GL barriers,  $I$ is the   IR radiation intensity (photon flux inside the heterostructure),  $d$ and $\kappa$ are the thickness of the inter-GL barrier layer and its dielectric constant, respectively, and $\delta (z)$ is the Dirac delta function (which reflects narrow localization of the electron and hole charges in the GLs),
In contrast to QWIPs using the intraband transitions,
the photoexcitation rate in GLIPs is virtually independent of $\Sigma_n^e$.
Both thermalized electron tunneling rate $G_n$ and 
 the escape probability of the photoexcited electrons $\theta_n$ from the $n$-th GL (and, hence, the photoemission rate) can markedly, namely, exponentially  increase with the electric field
in the barrier under this GL $E_{n+1} = (d \varphi/d z)|_{nd  \leq z \leq (n+1)d}$. 
The escape  probability of the photoexcited electrons
is also determined by the effective energy of the photoexcited electron $\varepsilon = \hbar\Omega/2 $.
Although  the  parameter $p_n$ characterizing the capture  into the $n$-th GL   depends, to some extent, on the local electric field, 
 this dependence is much weaker than the field-dependence
of the tunneling from the GLs. In reality,  $p_n$ is most likely to be a functional of the electric-field spatial distribution~\cite{26,27,36}. This is particularly true if the energy relaxation length $l_{\varepsilon} $ is larger or comparable with the heterostructure thickness $Nd$. The dependence $p_n = p$ versus $V$ is  smoothly  decreasing with $V$ (see,  the previous calculations~\cite{26}).
Hence,  in the simplest approach can be presented as a function of the average field (in fact, of the bias voltage) as assumed in the following. In line with this, we set $p_n = p$, where $p$ is a function of $V$.

The tunneling  rates of the thermalized and photoexcited electrons  (neglecting the intraband Drude absorption in the emitter GL due to relatively high energies of IR photons)  can be presented as

\begin{equation}\label{eq3}
G_n 
= \frac{j_{0}}{e}
\exp\biggl(-\frac{E_{tunn}}{E_{n+1}}\biggr),
\end{equation}  
 
 \begin{equation}\label{eq4}
 \theta_n  = 
 \frac{(1 - \beta)^{n}}{1  +  \displaystyle\frac{\tau_{esc}}{\tau_{relax}}\exp\biggl( \frac{\eta^{3/2}\,E_{tunn}}{E_{n+1}}\biggr)}.
\end{equation} 
Here 
$E_{tunn} = 4\sqrt{2m} \Delta^{3/2}/3e\hbar$ is the field characterizing the   tunneling through the triangular barrier top~\cite{40,41}
 (see also~\cite{22,23}), $m$ is the electron mass in the barrier material, $j_{0}$ is the maximum current density which can be extracted from an {\it undoped} GL ($\varepsilon_F = 0$) at a given temperature, $\tau_{esc}$ is the try-to-escape time,
$\tau_{relax}$ is the characteristic time of 
the photoexcited electrons energy relaxation, 
  and  $\eta = [(\Delta - \hbar\Omega/2)/\Delta]$ if $\hbar\Omega/2 < \Delta$ and $\eta = 0$ if $\hbar\Omega/2 \geq \Delta$. 
The factor $\eta$
accounts for the difference
in the tunneling transparency of the thermalized and photoexcited (with the energy 
$\hbar\Omega/2$) electrons. Assuming $\Delta = 0.4$~eV and $m = 0.28m_0$ ($m_0$ is the free electron mass), we obtain $E_{tunn} \simeq 910 $~V/$\mu$m. 

At $\tau_{esc} \ll \tau_{relax}$, the right-hand side of equation~(3) corresponds to $\theta_n\simeq 1$ (i.e., the majority of the photogenerated electrons leaves the GLs. In the opposite case $\tau_{esc} > \tau_{relax}$,
$w_n \simeq \displaystyle
\frac{\tau_{relax}}{\tau_{esc}}\exp\biggl( - \frac{\eta^{3/2}\,E_{tunn}}{E_{n+1}}\biggr) $.
The energy relaxation of photoexcited electrons in the GL-structures under consideration
can be associated with (i) the emission of optical phonons with the transition of an electron to a low energy state in the GL conduction band (at $\hbar\Omega/2 > \hbar\omega_0 $, where $\hbar\omega_0 \sim 0.2$~eV is the optical phonon energy), (ii) the electron recombination
assisted by the emission of optical phonon (when $\hbar\Omega/2 > \hbar\omega_0$, and (iii) the collisions of the photoexcited
electrons with the thermal electrons and holes (in a wide range of IR photon energies $\hbar\Omega$. According to the above, one can set
 
\begin{equation}\label{eq5}
\frac{1}{\tau_{relax}} \simeq \frac{1}{\tau_0}   + \frac{1}{\tau_{ee}}.
\end{equation} 
Here
$\tau_0$ is the time of spontaneous optical phonon emission in GLs and 
$\tau_{ee}$ is the time of the photoexcited electrons relaxation on the thermalized electrons (and holes) in the GLs. The latter is much shorter in both undoped and doped GLs (see, for example,~\cite{40}).  In the following we assume
that $\tau_{relax} \simeq \tau_{ee}$ considering only the mechanism (iii). Since $\tau_{ee}$ depends on the electron density, we distinguish
$\tau_{relax,E}$ in the emitter GL and $\tau_{relax}$ in the inner GLs. As demonstrated in the Appendix, $\tau_{relax,E} > \tau_{relax}$. Although the ratio $\tau_{esc}/\tau_{relax}$ can be estimated for
the heterostructures with different barrier layers, we will mainly considered is as a phenomenological parameter.

In equation~(3), we have neglected the thermionic emission from the GLs. The pertinent
contribution is proportional to the factor $\exp(-\Delta/T)$. The latter is assumed to be  small
compared with the factor $\exp(-E_{tunn}/E_{n+1})$ in equation~(3)
at the actual values of the electric field in the barriers and $T = 300$~K and lower.

Equation~(1) is supplemented by the following boundary conditions:

\begin{equation}\label{eq6}
\varphi|_{z = 0} = 0, \qquad \varphi|_{z = (N+1)d} = V, \qquad \sum_{n = 1}^{N+1}E_n = \frac{V}{d}.
\end{equation}

The current density $j$ is equal to the density of the  current injected from the emitter GL  through a pertinent barrier. 
Considering both the injection of the thermalized electrons from the doped emitter GL and the escape of the electrons photoexcited in this GL,
one can use the following formula relating  this current density and the electric field $E_1$:

\begin{eqnarray}\label{eq7}
j = j_{0}K\exp \biggl(-\frac{\gamma^{3/2}\,E_{tunn}}{E_1}\biggr)\nonumber\\
 +\frac{e\beta\,I}{1  +  \displaystyle\frac{\tau_{esc}}{\tau_{relax,E}}\exp\biggl( \frac{\eta^{3/2}\,E_{tunn}}{E_{1}}\biggr)}.
\end{eqnarray}
Here 
$\gamma  = [(\Delta- \mu_E)/\Delta]$ is the work function from the emitter to the barrier layer normalized by the conduction band offset. 
The factor $K \geq 1$ is associated with an increase of the thermalized electron density  by the electric field in the emitter barrier layer $E_1$ in the GLIPs with the -type emitter layer  (see the Appendix).  
One needs to stress that the value of $\gamma$ in the  GLIPs with the emitter GL of p-type
is larger whereas $K$ is smaller ($K< 1$) then those for 
the  GLIPs  with the n-type emitter GL (due to different signs of the Fermi energy counted with respect to the Dirac point)
  .

\section{Dark current characteristics}

\begin{figure}[t]
\centering
\includegraphics[width=7.0cm]{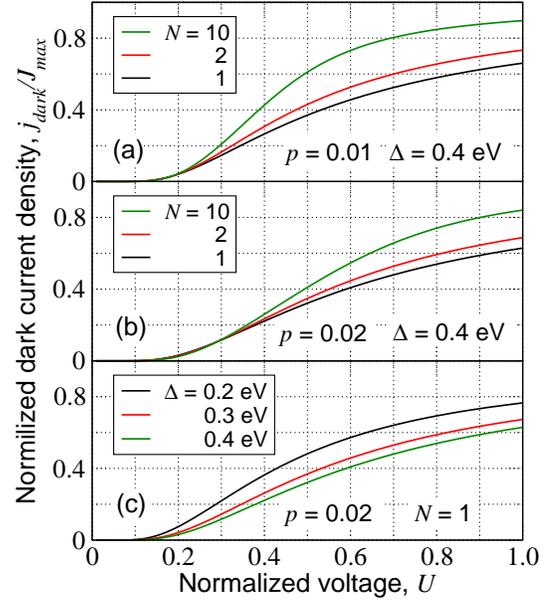}
\caption{Dark current-voltage characteristics $j_{dark}/J_{max}$ versus $U = V/(N+1)dE_{tunn}$ of GLIPs  with 
(a)different $N$ for   $\Delta = 0.4$~eV and $p = 0.01$, (b)different $N$ for  $\Delta = 0.4$~eV and $p = 0.02$,, and (c) different $\Delta$ at $N = 1$ and $p = 0.02$.
}
\end{figure}

In the absence of irradiation ($I_0 =0$), the current  across the GLIP constitutes the 
dark current, so that
$j = j_{dark}$. As follows from equations~(2), (3), and (7),the electric fields in the barriers are
the following functions of the dark current density: 

\begin{equation}\label{eq8}
E_{1}^{(dark)} = \frac{\gamma^{3/2}E_{tunn}}{\ln (j_{0}K/j_{dark})}=  E_E^{(dark)}
\end{equation}
in the near  emitter barrier ($n = 1$), and

\begin{equation}\label{eq9}
E_{n}^{(dark)} = \frac{E_{tunn}}{\ln (j_{0}/j_{dark}p)} = E_B^{(dark)}
\end{equation}
in other barriers ($n = 2,..., N+1$).
Taking into account conditions ~(6) and using equations (8) and (9), we obtain

\begin{equation}\label{eq10}
 \frac{V}{dE_{tunn}}=\frac{\gamma^{3/2}}{\ln (j_{0}K/j_{dark})} + \frac{N}{\ln (j_{0}/j_{dark}p)}.
\end{equation}
Equation (10) yields the following expression for the dark current-voltage characteristics

\begin{eqnarray}
j_{dark} = J_{max}\exp\biggl\{- \frac{dE_{tunn}}{2V}\biggl[
\biggl(\gamma^{3/2} + N  + \frac{V\ln\,Kp}{dE_{tunn}}\biggr) \nonumber\\
+\sqrt{\biggl(\gamma^{3/2} + N  + \frac{V\ln\,Kp}{dE_{tunn}}\biggr)^2 
- \frac{4\gamma^{3/2}V\ln\,Kp}{dE_{tunn}} 
}\biggr]
\biggr\},
\end{eqnarray}\label{eq11}
where $J_{max} = j_{0}K$.

   From equation~(11) we obtain

\begin{equation}
j_{dark} \simeq  \frac{J_{max}}{\Gamma}\,\exp\biggl[-\displaystyle\frac{(\gamma^{3/2} + N)dE_{tunn}}{V}\biggr].
\end{equation}\label{eq12}
Here the quantity
\begin{equation}\label{eq13} 
\Gamma = (Kp)\exp\biggl[-\frac{\gamma^{3/2}\ln Kp}{(\gamma^{3/2} + N +V\ln Kp/dE_{tunn})}\biggr]
\end{equation} 
changes from $\Gamma \simeq (Kp)^{N/(\gamma^{3/2} +N)}$ if $V \ll (\gamma^{3/2} +N)dE_{tunn}$  to
$\Gamma \simeq Kp$ at $V \gg (\gamma^{3/2} + N)dE_{tunn}$. Hence $\Gamma$ is rather smooth function of $V$.

Consequently, at moderate voltages
$V \ll (\gamma^{3/2} +N)dE_{tunn}$, 
corresponding to the most practically interesting conditions when the dark current is small,

\begin{eqnarray}
E_E^{dark} \simeq \frac{\gamma^{3/2}V}{(\gamma^{3/2} + N)d}\biggl[1 - \displaystyle\frac{NV\ln Kp}{(\gamma^{3/2} + N)^2dE_{tunn}}\biggr]\nonumber\\
\simeq\frac{\gamma^{3/2}V}{(\gamma^{3/2} + N)d}, 
\end{eqnarray} \label{eq14}

\begin{eqnarray}
  E_B^{dark} = \frac{V}{(\gamma^{3/2} + N)d}\biggl[1 + \displaystyle\frac{\gamma^{3/2}V\ln Kp}{(\gamma^{3/2} + N)^2dE_{tunn}}\biggr]\nonumber\\
\simeq \frac{V}{(\gamma^{3/2} + N)d}.
\end{eqnarray}\label{eq15}

It is instructive that, according to equation~(12), $j_{dark}$ being proportional 
to  $p^{-N/(\gamma^{3/2} +N)}$ strongly increases with decreasing capture efficiency $p$,
particularly if $\gamma^{3/2} \ll N$ (i.e., at  $N > 1$.  This implies that at sufficiently 
small $p$ when  $pN <1$, the net dark current can
 exceed the current of the tunneling thermalized electrons (which is approximately proportional to the number of inner GLs $N$), i.e., the main component of the dark current is associated with the current injected from the emitter controlled by
the tunneling from the internal GLs being, certainly, smaller than $J_{max}$. 
Another feature of the GLIP current-voltage characteristics and the electric-field spatial distributions
given by equations~(12) - (14) is their weak dependence on the parameters of the emitters $\gamma$ and $K$ (at
$\gamma^{3/2} < 1$ and particularly when $N \gg 1$).
The latter says that if $\gamma^{3/2} \ll N$,
the tunneling emitter under consideration can be considered as an ideal. 


%

One needs to stress that due to a decreasing voltage dependence of the capture efficiency $p$,
this dependence provides an increasing pre-exponential extra factor  in the dark current-voltage relation given by equation~(12)
and  enhances its  steepness. 

Considering the equality of the electric fields in all  layers except the emitter barrier layer
[see equations (8) and (9)] and using equations (13),  the charge density of the  GL adjacent to the emitter GL $Q_1 = -(\kappa/4\pi)(E_{B}^{(dark)} - E_{E}^{dark}) =
Q_{E}$  is given by

\begin{equation}\label{eq16}
Q_E
\simeq \frac{\kappa}{4\pi}\biggl(\frac{\gamma^{3/2} - 1}{\gamma^{3/2} +N}\biggr)\frac{V}{d}.
\end{equation} 
Depending on  the emitter GL doping (n- or p-type)), which determines $\gamma$ ($\gamma < 1$ and and $\gamma >1$ in the n- and p-type emitter GL, respectively ), the charge density
 $Q_E$ can be both negative 
(i.e., the electron density exceeds the hole density) and positive (the hole density is larger than the electron density),
whereas the charge densities of other inner GLs $Q_n = Q_B$ ($n = 2,3,...,N$) are equal to zero.

The case $\gamma < 1$ when $E_E^{(dark)} < E_B^{(dark)}$ and $Q_E < 0$, corresponds to the potential in the GLIP heterostructure profile schematically shown in figure~2(a).

Figure~3  shows examples of the normalized dark current density $j_{dark}/J_{max}$ versus the normalized voltage $U = V/(N+1)dE_{tunn}$  calculated using equation~(11)  for the GLIPs with  $\varepsilon_F = 0.1$~eV  and different  number of the inner GLs $N$ for different parameters  $\Delta$ and $p$.
The comparison of the plots in figures~3(a) and 3(b) confirms a weak dependence of the dark current on the number of GLs $N$  and a pronounced rise of the dark current with decreasing capture efficiency $p$ (particularly at lower normalized voltages $U$). Higher dark currents at smaller  band-offsets $\Delta$ are naturally attributed to   
tunneling emission strengthening from all the GLs with decreasing of the  tunneling barrier height  $\Delta$.


As follows from equation~(12) and figure~3, $j_{dark}$  drastically decreases with decreasing voltage. However, at small voltages,
$E_E^{dark}$ can be so small that the tunneling exponent $\exp(-\gamma^{3/2}E_{tunn}/E_E^{dark})$ becomes smaller than
the thermionic exponent $\exp(-\gamma\Delta/T)$~\cite{42}. This implies that at such voltages, the tunneling injection from the emitter GL  gives way to
the thermionic emission. This determines the minimal value of $V$, at which equations (11) - (13)  are still valid: $V > V_{min} \simeq dE_{tunn}(T/\gamma\Delta)(\gamma^{3/2} + N )$. 
At $V \sim V_{min}$,
$j_{dark} \sim (j_{0}/p)\exp(- \gamma\Delta/T)$. This implies that the dependences shown in figure~3
are correct up to small values of $U$ (at $U > 0.08 - 0.12$ for $\Delta = (0.2-0.4)$~eV)

\section{Photoresponse}

We limit our consideration of the GLIP operation primarily by the situations  when the incident IR radiation is  relatively weak.

In the case  of  irradiation, equations~(9) and (10) should be replaced by
more general equations accounting for  the tunneling emission of both the thermalized 
and photoexcited electrons. Since the photoexcitation leads to the variations of the electric fields in the barrier layers, 
 equation~(10) in this situation should be replaced by the following equation which governs the photocurrent density $j_{photo} = j - j_{dark}$:

\begin{eqnarray}\label{eq17}
\frac{V}{dE_{tunn}} =\frac{\gamma^{3/2}}{\ln [j_{0}K/(j_{dark} + j_{photo} - e\beta\,I\theta_0)]}\nonumber\\
 + \sum_{n =1}^{N}\frac{1}{\ln [j_{0}/(j_{dark}p + j_{photo}p - e\beta\,I\theta_n)]},
\end{eqnarray}
where $\theta_n$ is defined by equation (4) and $j_{dark}$ is governed by equation ~(10) leading to equation~(11) and its consequences.
At relatively low IR radiation intensities, such as $e\beta\,I \ll j_0$,  
 the electron photoexcitation can be treated 
as a perturbation.  
Considering this, into the quantities $\theta_0$ and $\theta_n$ in equation (17) on can insert the unperturbed values of the electric fields, namely, $E_E^{(dark)}$ and $E_B^{dark}$.   As a result, 
for the photocurrent density $j_{photo}$ from equation~(17)
we obtain

\begin{eqnarray}
j_{photo} = e\beta\,I\,\frac{\biggl[\gamma^{3/2} + \displaystyle\frac{(1-\beta)[1 - (1 -\beta)^{N}]}{p\beta}\biggr]}
{\biggl( \displaystyle\frac{\Theta_{dark}^2}{\gamma^{3/2}} + N\biggr)}\nonumber\\
\times \frac{1}{1  +  \displaystyle\frac{\tau_{esc}}{\tau_{relax}}
\exp\biggl( \frac{\eta^{3/2}\,E_{tunn}}{E_B^{(dark)}}\biggr)}.
\end{eqnarray}\label{eq18}
Here $\Theta_{dark} = (E_E^{(dark)}/E_B^{(dark)})^2$ characterizes the GLIP in dark.
According to equations (8) and (9), $\Theta_{dark}$ is expressed via the dark current density and, invoking equation (12), via the voltage:

\begin{eqnarray}\label{eq19}
\Theta_{dark} = \gamma^3 \biggl[\frac{\ln (J_{max}/j_{dark}Kp)}{\ln (J_{max}/j_{dark})}\biggr]^2\nonumber\\
= \gamma^3\biggl[1 +  \frac{V\,\ln(1/Kp)}{(\gamma^{3/2} + N)dE_{tunn}}\biggr]^2.
\end{eqnarray}
If $Kp = 1$ or at moderated voltages $V <
 (\gamma^{3/2} + N)dE_{tunn}$, one can put in equation~(18)
$\Theta_{dark} = \gamma^3$,
so that  one obtains

\begin{eqnarray}
j_{photo} = \frac{e\beta\,I}{(\gamma^{3/2} + N)}\nonumber\\
\times\frac{\biggl[\gamma^{3/2} +   \displaystyle\frac{(1 - \beta)[1 - (1 -\beta)^{N}]}{p\beta}\biggr]}{1  +  \displaystyle\frac{\tau_{esc}}{\tau_{relax}}\exp\biggl[ \frac{\eta^{3/2}(\gamma^{3/2} + N )\,dE_{tunn}}{V}\biggr]}
\end{eqnarray}\label{eq20}
The appearance of $\gamma^{3/2}$ in  equation~(20) is associated with the contribution 
 of the photoexcitation from the emitter GL. Since  normally $\gamma^{3/2}p \ll 1$, for the GLIPs with not too large GLs ($N < \beta^{-1} \sim 40$), equation~(20) can be simplified and presented in the following form: 

\begin{equation}
j_{photo}\simeq
\frac{\displaystyle\frac{e\beta\,I}{p}\frac{N}{(\gamma^{3/2} + N)}}{1  +  \displaystyle\frac{\tau_{esc}}{\tau_{relax}}\exp\biggl[ \frac{\eta^{3/2}(\gamma^{3/2} + N )\,dE_{tunn}}{V}\biggr]}
.
\end{equation}\label{eq21}

It is worth noting   that the photocurrent density given by equations (18), (20), and (21) exceeds the current created by the electrons photoexcited from the emitter GL by a factor $p^{-1}$, which can be very large.
This is because the photoemission from the inner GLs stimulates a strong injection of the thermalized electrons from the emitter GLs resulting in a substantial amplification of the photocurrent.

Taking into account equation (20), one can find small variations of the electric fields  caused by the irradiation in all barrier layers
$E_n - E_n^{(dark})$ as functions of the bias voltage $V$ and the radiation intensity $I$. In particular, 
the electric field in the emitter barrier layer $E_E = E_1 < E_2$ (as under the dark conditions). However, due to the attenuation of the IR radiation as its propagates across the GLIP structure,
 the electric-fields in the consequent GLs slightly increase: $E_2 < E_3, E_3 <E_4,...$ 
 (such a change is proportional to the radiation intensity $I$,
which is weak in the case under consideration).

The photoexcitation leads to the variations of the electric fields in the barrier layers: an increase in $E_E$ and  in a decrease in $E_B$.
At sufficiently strong irradiation, the electric field in the emitter barrier $E_E$ can become larger than the electric field in the bulk $E_B$.  In this case, the GL adjacent to the emitter GL is charged positively, and the potential profile 
becomes similar to that shown schematically in figure~2(b). 
An  high radiation intensities, the electric field in the heterostructure bulk drops, so that the potential
drop across the emitter layer tends to the applied voltage $V$. This implies that $E_E$ tends to $V/d$. Consequently,
the net current density tend to a value which roughly can be estimated as $j \simeq J_{max}\exp(-\gamma^{3/2}dE_{tunn}/V) + e\beta\,I[1 + (\tau_{esc}/\tau_{relax}\exp(\eta\,dE_{tunn}/V))]^{-1}$.
However 
at very high radiation intensities, the GLIP operation might, in particular,  be complicated by the following effects: the saturation of the interband absorption~\cite{43},
the influence  of  elevated density of the photoexcited electrons on their escape~\cite{23,44}, and the influence of  the space-charge of the electrons propagating across the barrier layers on the electric-field distributions. These  effects are beyond of our model framework.

\section{Responsivity and other GLIP characteristics}

\begin{figure}[t]
\centering
\includegraphics[width=7.0cm]{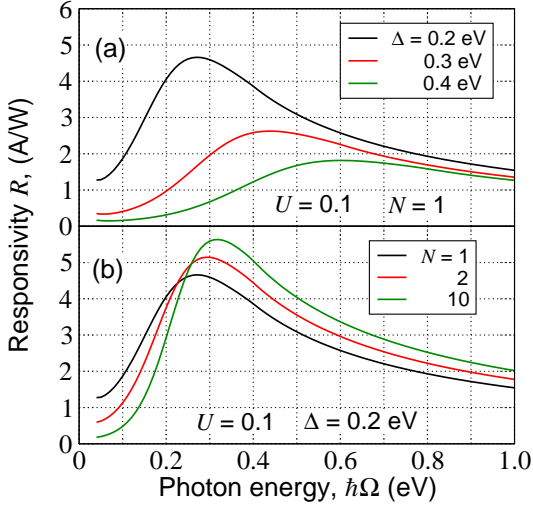}
\caption {Spectral characteristics of GLIPs with (a) $N = 1$ and different $\Delta$
and (b) $\Delta = 0.2$~eV and different $N$ at $U = 0.1$ ($\tau_{esc}/\tau_{relax} = 0.1$, and $p =0.01$)}
\end{figure}

\begin{figure}[t]
\centering
\includegraphics[width=7.0cm]{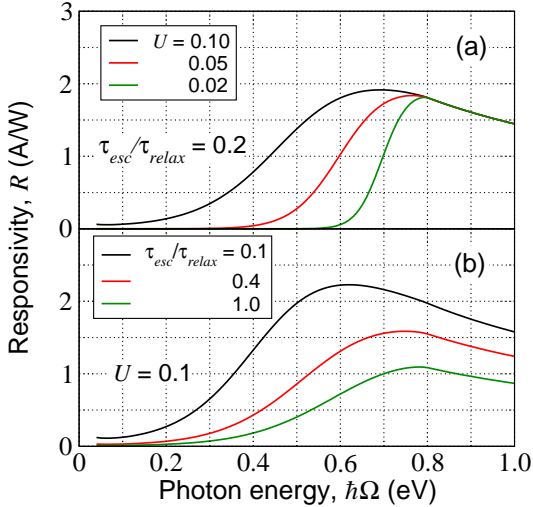}
\caption {The same as in figure~4, but    (a) for  $\tau_{esc}/\tau_{relax}
= 0.2$ at different normalized voltages $U$ and (b) for different ratios $\tau_{esc}/\tau_{relax}$ at $U = 0.1$ ($\Delta = 0.4$~eV, $N = 2$, and
$p = 0.02$).}
\end{figure}

The detector responsivity ${\cal R}$ is usually defined as

\begin{equation}\label{eq22}
{\cal R} = \frac{Aj^{photo}}{S}.
\end{equation}
Here $A$ is the device area, $S = A{\cal I}\hbar\Omega$ is the incident IR radiation power, and ${\cal I}$ is the incident IR radiation flux (hence, $I = {\cal I}(1-r)$, where $r$ is the coefficient 
 of 
 the IR radiation reflection from the GLIP top.
Considering this and using equation~(21), we arrive at the following formula for the 
GLIP responsivity:

\begin{equation}\label{eq23}
{\cal R} \simeq \overline{{\cal R}}\frac{\displaystyle\frac{N}{(\gamma^{3/2} + N)}}
{1 + \displaystyle\frac{\tau_{esc}}{\tau_{relax}}
\exp\biggl[\frac{\eta^{3/2}(\gamma^{3/2}+ N)dE_{tunn}}{V}\biggr]
}.
\end{equation}
Here ${\overline{\cal  R}} = {\overline R}\xi$, where

\begin{equation}\label{eq24}
{\overline{\cal  R}} = \frac{e\pi\alpha}{\hbar\Omega\,p}
\end{equation}
is the characteristic responsivity and the factor $\xi = (1-r)/\sqrt{\epsilon}$ is determined by the conditions of reflection
of the incident IR radiation from the GLIP top interface (the presence or absence of an anti-reflection coating) and its  propagation across the heterostructure. The reflection from the substrate can also affect the value of $\xi$. In the simplest case, the reflection factor $\xi$
can be expressed via $\sqrt{\epsilon}$: $\xi = 4/(\sqrt{\epsilon} + 1)^2$.

The deviation of the nominator in equation (23) from unity reflects the dependence of the responsivity
on the emitter parameter $\gamma^{3/2}$. This  deviation practically vanishes at large $N$ and $\gamma^{3/2} < 1$.
In the latter limit, the emitter can be considered as an ideal (i.e., providing the necessary injection to compensate the photoexcitation from the inner GLs~\cite{24,30,31}), and the GLIP responsivity is practically independent
of $N$ if the average electric field $V/Nd$ is fixed.

Since in GLIPs the  coefficient $\pi\alpha$ is relatively large
and the capture efficiency $p$ is relatively small ($p \simeq 0.5 \%$
~\cite{28}) compered to the standard heterostructures, the characteristic 
responsivity ${\overline R}$ can be fairly large. Indeed, setting $\hbar\Omega = 0.1 - 1.0$~eV and $p = 0.01$, we obtain ${\overline R} \simeq (2.3 - 23.0)$~A/W

Figures 4 and 5 show the responsivity, $R$ (normalized by the reflection factor $\xi$)  spectral characteristics of the GLIPs 
with different band offsets $\Delta$,
 number of the barrier layers $N$, normalized   voltages $U = V/(N+1)dE_{tunn}$, capture efficiency $p$,
 and $\tau_{esc}/\tau_{relax}$. 
In these and the next figure  it is assumed that
the electron Fermi energy in the emitter GL $\varepsilon_F = 0.1$~eV. 
Figure~6 shows the responsivity $R$ as a function of the normalized voltage calculated for the GLIPs with different
$\Delta$.

Among the main features of the responsivity spectral dependence, the following should be pointed out:
(a) Weak dependence of the spectral characteristics on thee number of the GLs $N$;
(b) An increase in the responsivity with decreasing capture efficiency $p$;
(c) A marked  shift of the responsivity maximum toward larger photon energies with increasing the band-offset $\Delta$,
normalized voltage $U$, and relative escape time $\tau_{esc}/\tau_{relax}$;
(d) The responsivity maxima correspond to the photon energies markedly smaller than $\hbar\Omega = 2\Delta$,
so that the GLIPs can exhibit relatively high response at $\hbar\Omega < 2\Delta$.
A fairly pronounced dependence of $R$ on $U$ (i.e., the applied voltage) implies the possibility of an effective 
voltage control

It is important to stress that at small values of the capture parameter $p$,
the photocurrent can markedly exceed the current created solely by
the electrons photoemitted from the GLs due to the effect of photoelectric gain,
which implies that the electron photoemission stimulates much stronger tunneling emission 
from the emitter.

Taking into account that  the net photoemission rate  from all GLs is equal to 

\begin{equation}\label{eq25}
G_{photo} \simeq \frac{ N\beta\,I}{1 + \displaystyle\frac{\tau_{esc}}{\tau_{relax}}\exp\biggl[
\frac{\eta^{3/2}(\gamma^{3/2} + N)dE_{tunn}}{V}\biggr]
},
\end{equation}
the photocurrent density (21) )can be presented as 

\begin{equation}\label{eq26}
j_{photo} = eG_{photo}gI,
\end{equation}
where
\begin{equation}\label{eq27}
g = \frac{1}{p(\gamma^{3/2} +N)}
\end{equation}
 is the photoelectric gain. Due to  small values of $p$ in GLIPs, $g \gg 1$ even at fairly
  large (but realistic) $N$.

The dark current limited detector detectivity, ${\cal D}^*$, and its normalized version $D^* = {\cal D}^*/\xi$ are expressed via the responsivity and the dark current
density as (for example,~\cite{24}):

\begin{equation}\label{eq28}
{\cal D}^* = \frac{{\cal R}}{\sqrt{4egj_{dark}}}, \qquad D^* = \frac{R}{\sqrt{4egj_{dark}}}.
\end{equation}
Here for $j_{dark}$, $R$, and $G$ one can use equations (12), (23), and (25), respectively.
Considering this, we find (at $N < \beta^{-1}$)

\begin{eqnarray}
D^* = \overline{D^*}\frac{N}{\sqrt{(\gamma^{3/2} + N)}}
\nonumber\\
\times\frac{\displaystyle\exp\biggl[\frac{(\gamma^{3/2} +N)dE_{tunn}}{2V}\biggr]}{1 + \displaystyle\frac{\tau_{esc}}{\tau_{relax}}\exp\biggl(
\frac{N\eta^{3/2}dE_{tunn}}{V}\biggr)}
\label{eq29}
\end{eqnarray}
Here  
\begin{equation}
\overline{D^*} =\frac{e\beta}{\hbar\Omega\sqrt{4ej_{0}}\,\displaystyle (Kp)^{\gamma^{3/2}/(\gamma^{3/2}+N) }}.
\end{equation}\label{eq30}
Equation~(29) immediately shows that the GLIP detectivity increases with the increasing number
of the inner GLs $N$, while the responsivity (see equation~(23)) 
is a weak funcion of $N$.
Indeed,  at large $N$, $D^* \propto \displaystyle\frac{N}{\sqrt{(\gamma^{3/2} + N)}} \simeq \sqrt{N}$, but
$R$ is virtually independent of $N$. Some deviation from the latter is associated with the emitter effect
(characterized by $\gamma^{3/2}$). The spectral dependence of $D^*$ repeats that of $R$ (see figures~4 and 5).

The detectivity $D^*$ markedly increases with the decreasing voltage $V$. 
This is owing to the pertinent drop of the dark current.
However, since 
at  small values of $V$, the tunneling current from the emitter GL
can become smaller than the thermionic current, substituting $V_{min}$ (see above) to equation~(30), we obtain the following estimate for the maximum of $D^*$: 

\begin{equation}
D _{max}^* \simeq  \frac{\overline{D^*} \displaystyle\exp\biggl(\frac{\gamma\Delta}{2T}\biggr)}{1 + \displaystyle\frac{\tau_{esc}}{\tau_{relax}}\exp\biggl[
\frac{N}{(\gamma^{3/2} + N )}\frac{\eta^{3/2}\gamma\Delta}{T}\biggr]}.
\end{equation}\label{eq31}
At $\hbar\Omega \simeq 2\Delta$, the parameter $\eta$ can be small. In this case,
equation (31) yields

\begin{equation}
D _{max}^* \simeq \frac{\overline{D^*}\exp (\gamma\hbar\Omega/T)}{(1 + \tau_{esc}/\tau_{relax})} \propto \frac{\beta\exp (\gamma\hbar\Omega/T)}{\hbar\Omega(1 + \tau_{esc}/\tau_{relax})}.
\end{equation}\label{eq32}

\begin{figure}[t]
\centering
\includegraphics[width=7.0cm]{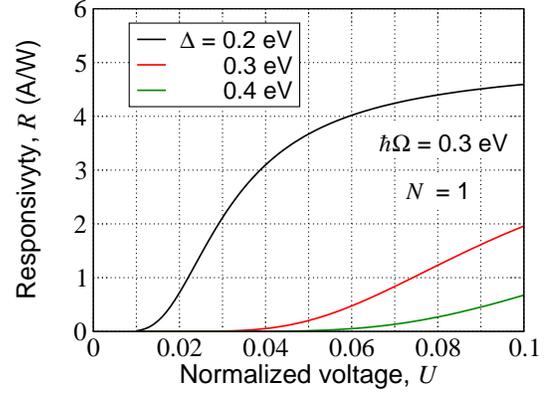}
\caption {Voltage dependences of   the  responsivity
of GLIPs with $N = 1$ and differerent $\Delta$ for $\hbar\Omega = 0.3$eV, $\tau_{esc}/\tau_{relax} = 0.1$, 
and $p =0.01$.
}
\end{figure}

\section{Comparison with  other IR photodetectors}

The comparison of GLIPs and QWIPs (based, in particular, on InGaAs/AlGaAs and other heterostructures with the n-type QWs)  intended for the operation in mid- and near IR ranges,
shows that GLIPs can exhibit the following advantages: (i) sensitivity to normally incident radiation (no radiations coupling structures are needed; (ii) higher probability of the photo-assisted escape from the GL  bound states to  the continuum states above the barriers due to a relatively large universal constant $\pi\,e^2/\hbar\,c \simeq 2.3\%$.
(compare with $\beta \sim 0.5 \%$ for a GaAs QW with the typical electron density $~\sim 5\times 10^{11}$~cm$^{-2}$~\cite{24}); smaller  capture efficiency of the electrons propagating over the barriers  the into the GLs (up to the orders of magnitude - compare the data from~\cite{28} and~\cite{30}). 
These advantages imply that the GLIP responsivity can exceed that of QWIPs. 
 Since the modulation bandwidths of both GLIP and QWIPs
is determined by the vertical electron transport across the barrier layers, one might assume that these bandwidths are can close to each other.

As follows from equation~(21) and equation~(4) from~\cite{8}, the  responsivity, $R$,  of a GLIP and the responsivity,
$R_{GLPD}$ of a lateral
GL-photodiode (GLPD) with a lateral p-i-n junction ,
are (at $\tau_{esc} \ll \tau_{relax}$) approximately equal to 

\begin{equation}\label{eq33}
R 
\simeq \frac{e\beta}{\hbar\Omega\,p}, \qquad R_{GLPD} \simeq\frac{2e\beta\,N}{\hbar\Omega},
\end{equation}
respectively. Hence, 

\begin{equation}\label{eq34}
\frac{R}{R_{GLPD}} \simeq \frac{1}{2pN}.
\end{equation}
Since, the capture efficiency $p$ can be very small, the above ratio can exceed unity. 

Since the electron transit time in the GLIPs (which are vertical transport devices)) can be shorter than in the GLPDs
(with the lateral transport), the GLIPs can surpass the GLPDs in speed.

In the double-GL GLIPs~\cite{22,23} and the UTC-PDs based on InAs/InP heterostructures~\cite{45,46}, the electrons photoexcited in the  emitter transfer (via the tunneling through the triangular barrier) to the undoped drift region. 
In contrast to  the  double-GL GLIPs and UTC-PDs, in which the responsivity is determined by the escape of the electrons photogenerated in the emitter layer,  the GLIPs can exhibit the effect of amplification of the current produced the electrons photoexcited  from the inner GLs. This effect in the GLIPs works
at a weak electron capture into the inner GLs, i.e.,  if $p \ll 1$, but at moderate modulation frequencies. 
The GLIP, double-GL GLIP, and  UTC-PD response  at relatively high modulation frequency  
 is affected by
the  electron transit  across this region. One can assume that  the electron transit  time in UTC-PDs can be somewhat shorter than in the GLIPs because of better transport properties of the drift region materials in the former devices.
However, the delay in the escape of the photoexcited electrons from the emitter limited by their diffusion (or in the best case, by the drift in the built-in electric field) across the heavily doped emitter  layer, can be the main factor
limiting the UTC-PD modulation bandwidths. This implies that the GLIPs (without such a delay in the emitter GL) can exhibit a high-speed performance close to that of the UTC-PDs and, hence, compete with the latter.  

Similar IR photodetectors can be based on the HgCdTe heterostructures with multiple zero-gap 
QWs, although the specifics of these heterostructures requires a separate study. 

\section{Conclusions}
We proposed the 
GLIPs - infrared photodetectors based  on vdW hetereostructures  with GLs  serving as the emitter, collector and absorbing layers.
The GLIPs exploit the interband electron transitions between the valence band in the GLs and the continuum states in the conduction band of the barrier materials. The GLIPs should be able to
operate in  different ranges  of the IR spectrum depending on the values of the conduction band offset between the GL and the barrier layer material. 
We showed that due to small capture probability of the excited electrons into the GLs together with a relatively strong interband photon absorption, the GLIPs 
can exhibit elevated photoelectric gain and responsivity.
 An increase in the number of  GLs in the GLIP heterostructure can provide relatively high values of the GLIP detectivity. Among other GLIP advantages are highly
conducting transparent contact GL layer, sensitivity to normally incident radiation and  high speed operation. 
The GLIPs, which  add the diversity to the IR and THz detector technology, can surpass other already realized and newly proposed  photodetectors for the imaging and optical communication systems and THz photomixers. 

\section*{Appendix}

\subsection*{A1. Escape frequency and relaxation time}

The tunneling rates of the thermalized and photoexcited electrons
depends on the electric field, $E$, at the GL. In equation~(4) we set the escape frequency equal to $\nu_{esc} = \exp(- E_{tunn}/E)/\tau_{esc}$ and $\nu_{esc} = 
\exp(- \eta^{3/2}E_{tunn}/E)/\tau_{esc}$ for the thermalized and photoexcited electrons respectively. Thus, it was assumed
that $\nu_{esc}$ is an exponential function of  $\eta^{3/2}E_{tunn}/E$.

 For more accurate calculations of the escape frequency $\nu_{esc}$ as a function of 
the quantity $\eta$ we found the Schrodinger equation  solutions with the complex energy for GLs embedded in 
a dielectric in the transverse electric field. The GL potential was modeled by a one-dimensional delta-well
~\cite{26}: $U(z) = \delta(z)\hbar^2/2ml(p)$, where $\delta(z)$ is the Dirac delta-function and $l(p)  = \hbar/2\sqrt{2m(\Delta - pv_W)}$ is the electron wave function localization length. This quantity depends on the in-plane electron momentum due to the mismatch of dispersion laws in the  GL and the barrier layer~\cite{23,47}.
Figure~7 shows the  $\nu_{esc}$ vs $\eta$ dependences calculated for different values of electric field $E$
using the above approach. The related dependences
calculated using the above simplified formula are also shown by dashed lines for the material parameters corresponding to WS$_2$ barrier layers ($\Delta = 0.4$~eV, $m = 0.28m_0$, and $E_{tunn} = 910$~V/$\mu$m).

At small $\eta$ the fit is not so ideal but can be considered as reasonable somewhat affecting   the shape of the GLIP spectral characteristics 
only at $\hbar\Omega \simeq 2\Delta$.

The inverse relaxation time  assumed to be primarily determined by the electron-electron scattering, so $1/\tau_{relax} = 1/\tau_{ee}$ as functions of  $\eta$ in the undoped GLs ($\varepsilon_F = 0$) and doped emitter GL ($\varepsilon_F = 100$~meV) at $\kappa = 5$ and $T = 300$~K are  GLs shown in figure~7 by the dotted lines. These curves were obtained by calculating the imaginary part of the electron RPA self-energy~\cite{44}.
 As follows from figure~7, in the most interesting actual range of 
$\eta < 0.2- 0.3$ the order-of magnitude estimates yields $\tau_{relax} \sim 3\times 10^{-14}$~s and 
$\tau_{relax,E} \sim  5\times 10^{-14}$. The latter characteristic times are markedly smaller that
the characteristic time of the spontaneous emission of optical phonons $\tau_0$ for both the intraband and interband electron transitions. Thus, for GLIPs with WS$_2$ barriers one can arrive at the following rough estimate:  $\tau_{esc}/\tau_{relax} \sim 0.10 - 0.17$.

\begin{figure}[t]
\centering
\includegraphics[width=7.5cm]{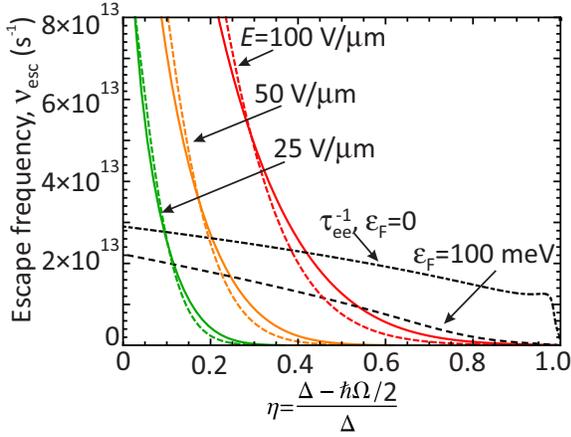}
\caption{Escape frequency $\nu_{esc}$ as a function of the normalized photon energy $\eta$
for different electric fields $E$ calculated using  a rigorous model (solid lines) and  a simplified model with constant pre-exponential factor (dashed lines). The dotted lines correspond to the  inverse relaxation time 
$\tau_{relax}^{-1} \simeq \tau_{ee}^{-1}$ associated with the scattering on the thermalized electrons (and holes) for the undoped and doped GLs ($\varepsilon_F = 0$~meV  and ($\varepsilon_F = 100$~meV, respectively) at $T = 300$~K.
}
\end{figure}

\subsection*{A2. Field-dependence of the capture efficiency}

Neglecting above the local dependence of the capture parameter,
we found that the electric field distribution across the device  structure 
includes a low-field domain (in the near emitter barrier) with $E_1 = E_{E}$
and a high=field domain occupying the rest of the structure (bulk) where $E_n = E_{B}$
(see equations (8) and (9). However, if $p_n$ depends on $E_{n-1}$,
the near emitter domain might be extended to a few barriers~\cite{29}. 

It is worth mention that if the capture efficiency $p_n$ is a sufficiently strong function
of $E_{n-1}$, the monotonic electric-field spatial distributions can become unstable against
the perturbation with the length $2d$~\cite{36,37,38}. Such an instability can lead
to quasi-chaotic spatio-temporal electric field variations, which eventually result in the formation of stable quasi-periodic electric-field distributions~\cite{48}.


\subsection*{A3. Field-induced electrons in the emitter GL}

The electric field in the emitter barrier induces extra electrons in the emitter GL and,
hence, leads to an enhancement in the electron  Fermi level or the hole Fermi  level if this GL is donor or acceptor doped,
respectively.
As a result, the parameter $\gamma^{3/2}$ can markedly depend on $E_E$ (decreasing with increasing $E_E$) somewhat affecting
the current-voltage characteristics. Considering this effect, we find

\begin{equation}\label{eq35}
\varepsilon_{F,E} = \varepsilon_F\sqrt{(1 + E_{E}/E_{ind})},
\end{equation}
$E_{ind} = 4e\varepsilon_F^2/\kappa\hbar^2v_W^2$. Hence 

\begin{equation}\label{eq36}
\gamma_E^{3/2} = \biggl[\frac{\Delta_E -\varepsilon_{F}\sqrt{(1 + E_E/E_{ind}}}{\Delta}\biggr]^{3/2}.
 \end{equation}
Setting $\varepsilon_F = 0.2$~eV and $\kappa = 5$, one obtains $E_{ind} \simeq 1.18\times 10^2$~V/$\mu$m.  

If $E_E \ll E_{ind}$, $\gamma_E^{3/2} \simeq \gamma^{3/2}(1 - 3E_E/E_{ind})$.
In this case, $\exp(-\gamma_E^{3/2}E_{tunn}/E_E) \simeq K\exp(-\gamma^{3/2}E_{tunn}/E_{E})$, where 

\begin{eqnarray}
K = \exp\biggl(\frac{3\gamma_0^{3/2}\varepsilon_FE_{tunn}}{4(\Delta_E -\varepsilon_F)E_{ind}}\biggr)\nonumber\\
=
\exp\biggl[\frac{\sqrt{2m(\Delta_E - \varepsilon_F)}}{4\varepsilon_F}\frac{\kappa\hbar\,v_W^2}{e^2}
\biggr]> 1.
\end{eqnarray}\label{eq37}
At $(\Delta_E - \varepsilon_F) = 0.2 - 0.3$eV,  $\varepsilon_F = 0.1 - 0.2$~eV, $m =0.28m_0$, and $\kappa = 5$
one obtains $K \simeq 9 -14$.

Thus, at not too large voltages and emitter electric fields, the injection (from the contact) of the extra electrons in the emitter GL does not affect the shape of the current-voltage characteristics,
but leads to a replacement   of the quantity $j_{0}$ by $j_0K$. If the emitter barrier layer has the thickness 
$d_E$
different from $d$,  equation (6) should be  modified.

\subsection*{A4. Carrier heating in the emitter GL}

The IR irradiation can somewhat reinforce the thermionic electron emission from the emitter
GL  due to heating high-density 2D electron plasma in the doped emitter
GL (see~\cite{21,22,23}; this effect was disregarded in Section 5). 
The electron photoexcitation can lead to a heating of the carriers in the emitter GL slightly modifying the injection properties of the emitter. However, in the regimes considered above, the emitter properties weakly affect the GLIP characteristics, particularly, for the multiple GL devices~($N\gg 1$). 
This  mechanism as well as the photoemission from the emitter GL can be crucial for
the double-GL GLIPs without the inner GLs (i.e., with N=0)~\cite{22,23}. 
The operation principle of the double-GL GLIPs is similar to that used in the uni-travelling-carrier photodiodes (UTC-PDs)~\cite{46,47}. 
But such GLIPs, being potentially interesting, do not exhibit the effect of photoelectric gain, which can substantially increase the responsivity.

The bolometric mechanism associated with the electron heating due to the  Drude absorption can be efficient 
at the   relatively low radiation frequencies in the GL devices with  all of the GLs doped
  ~\cite{21} (not considered here).

\subsection*{A.5 Dynamic response}

The evaluation of the GLIP dynamic response to transient IR radiation requires a substantial generalization of our model. It will be done elsewhere. Here we refrain to a  qualitative reasoning. The dynamic operation of the GLIPs under consideration is determined by following  characteristic times:   the electron transit time across the GLIP $t_{trans}$, the time of the GLs charging by the injected current due to
the electron capture $t_{charge}$ and the time of emptying of the GLs by the electron photoexcitation 
from the GLs $t_{ex}$.

The transit time can be estimated as $t_{trans} \simeq (N + 1)d/<v>$, where $<v>$ is the average electron velocity
across the barrier layers. The latter can be set to be equal to the electron saturation velocity $v_s$.
The characteristic times $t_{charge}$ and $t_{ex}$ decrease with an increasing dc component of the incident IR radiation intensity. At small capture parameters, even at fairly strong irradiation, $t_{charge}, t_{ex} \gg t_{transit}$.

At the modulation frequencies of the incident IR radiation $\omega < 1/t_{charge}, 1/t_{ex}$, the GLIP responsivity
is well described by the above formulas [e.g., by equation~(28)]. In this modulation frequency range,
the responsivity is about $R \sim {\overline R} = (e\beta/\hbar\Omega\,p) \propto 1/p$ and virtually independent of the number of the GLs $N$.
However, in the range $1/t_
{charge}, 1/t_{ex} < \omega < 1/t_{trans}$, the electron capture does not manage to provide
the GL recharging. As a result, the electric field $E_E$ and the current injected from the emitter GL do not follow the temporal variations of the IR radiation. This implies that
 the mechanism of the photoelectric gain is effectively switch-off. This leads to smaller values of the responsivity,
which dependence on both $p$ and $N$ is rather complex. At higher modulation frequencies $\omega > 1/\tau_{trans}$, the responsivity markedly drops with increasing frequency
 $\omega$. The electron transit time $t_{trans}$ limits the GLIP modulation bandwidths. Setting $v_s \sim 10^7$~cm/s
 and $(N+1)d = 10 - 50$~nm, we obtain $f_{3dB} \sim 0.3 - 1.5$~THz.
The effect of the structural parameters on the dependences of the responsivity of single- and multiple-QWIPs on the radiation modulation frequency was studied previously~\cite{49,50}.
However, the results of these studies  can not be directly applied to GLIPs. Recent experimental studies of the high-speed operation of the photodetectors based on double-GL vdW heterostructures with the interband photoexcitation~\cite{51} appears to be fairly promising.

\section{Acknowledgments}
The authors grateful to V Aleshkin ,  A Dubinov and A Satou for useful discussion and to N Ryabova and S Boubanga-Tombet for assistance.
The work at RIEC  and UoA was supported by the Japan Society for Promotion of Science
(KAKENHI Grants No.23000008 and  No.16H06361).  V R, D S and V~L acknowledge the support by the Russian Scientific Foundation (Grants No.14-29-00277 and   No.16-19-10557) and the Ministry of Education and Science of the Russian Federation (Grant No.16.19.2014/K). 
The work at RPI was supported by the US Army Research Laboratory Cooperative
Research Agreement.


\end{document}